\documentclass[12pt,a4paper,aps,preprint,superscriptaddress,nofootinbib]{WileyMSP-template}
\usepackage[utf8]{inputenc}
\usepackage{graphicx}
\usepackage{amssymb}
\usepackage{textcomp}
\usepackage{amsmath}
\usepackage{tabularx}
\usepackage{bm}
\usepackage{times}
\usepackage{color}
\usepackage{slashed}
\usepackage{multirow}
\usepackage{verbatim}
\usepackage{cancel}
\usepackage{subfigure}
\usepackage{epstopdf}
\usepackage{mathrsfs}
\usepackage{geometry}
\usepackage{graphicx}
\usepackage[numbers,sort&compress]{natbib} 
\setcitestyle{numbers,open={},close={}}  
\usepackage[none]{hyphenat}
\usepackage{appendix}
\usepackage{ragged2e}
\usepackage{url} 
\usepackage{xspace}

\usepackage[colorlinks=true, pdfstartview=FitV, linkcolor=blue, citecolor=blue, urlcolor=blue]{hyperref}
\allowdisplaybreaks[4]

\begin{document}
	\sloppy
	\justify
	\pagestyle{fancy}
	\rhead{\includegraphics[width=2.5cm]{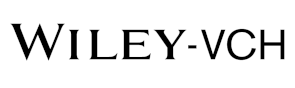}}

	\title{Detecting Axion Dark Matter by Artificial Magnetoelectric Materials}
	
	\maketitle

	% Author: Please give full first and last names for authors and include * after the name of all corresponding authors
	\author{Runyu Lei}
	\author{Chen-Hui Xie}
	\author{Jiayi Liu}
	\author{Zhong Liu}
	\author{Xin Liu}
	\author{Yu Gao*}
	\author{Sichun Sun*}
	\author{Jinxing Zhang*}

	% Dedication
	
	%\dedication{Optional dedication here. If no dedication is required, please leave blank}

	% Affiliations: Please provide adacemic titles (Prof. or Dr.) for all authors where applicable, and include an institutional email address for all corresponding authors
	\begin{affiliations}
		Runyu Lei\\%, Jiayi Liu, Zhong Liu, Xin Liu\\
		School of Physics and Astronomy, Beijing Normal University, Beijing 100875, China\\ Key Laboratory of Multiscale Spin Physics, Ministry of Education, Beijing Normal University, Beijing 100875, China
		
		Chen-Hui Xie\\
		School of Physics, Beijing Institute of Technology, Beijing 100081, China
		
		Jiayi Liu\\
		School of Physics, Beijing Institute of Technology, Beijing 100081, China
		
		Zhong Liu\\
		School of Physics and Astronomy, Beijing Normal University, Beijing 100875, China\\ Key Laboratory of Multiscale Spin Physics, Ministry of Education, Beijing Normal University, Beijing 100875, China
		
		Xin Liu\\
		Institute of Solid State Physics, Vienna University of Technology, Vienna 1040, Austria
		
		Yu Gao\\
		Institute of High Energy Physics, Chinese Academy of Sciences, Beijing 100049, China
		
		Sichun Sun\\ 
		School of Physics, Beijing Institute of Technology, Beijing 100081, China
		
		Jinxing Zhang\\
		School of Physics and Astronomy, Beijing Normal University, Beijing 100875, China\\ Key Laboratory of Multiscale Spin Physics, Ministry of Education, Beijing Normal University, Beijing 100875, China\\
		
		Email Address: gaoyu@ihep.ac.cn, sichunssun@bit.edu.cn, jxzhang@bnu.edu.cn
	\end{affiliations}
	
	% Keywords: Please provide a minimum of three and a maximum of seven keywords, separated by commas
	

	% Abstract should be written in the present tense and impersonal style (i.e., avoid we), and be at most 200 words long
	\begin{abstract}
		
		Axions are considered a key component of dark matter, characterized by very weak couplings to fermions and Chern-Simons couplings to gauge fields. We propose a novel detection mechanism based on symmetry-breaking magnetoelectric materials with a linear axionic coupling between magnetization and ferroelectric polarization. The focus is on a strain gradient  $\text{Sr}_2\text{IrO}_4$ film, where the breaking of space-inversion symmetry results in an emergent polar phase and an out-of-plane magnetic moment, exhibiting a flexomagnetoelectric effect. In this material, the linear $\mathbf{P} || \mathbf{M}$ enables a direct coupling between the external axion field and the intrinsic axion-like field, which amplifies the weak electromagnetic signals induced by axions, paving the way for pioneering axion detection. In contrast to conventional detection techniques, this mechanism is expected to enhance the sensitivity of the axion-electron and axion-photon coupling, providing a novel platform for axion detection and advancing the study of dark matter through the magnetoelectric effect.
	\end{abstract}
	
	% Text: Please use section headings and subheadings as specified below. For communications, all section headings apart from Experimental Section should be removed
	% Please make the first reference to a display item bold: \textbf{Figure 1}
	% Do not abbreviate Figure, Equation, etc.; display items are always singular, i.e., Figure 1 and 2.
	% Equations are always singular, i.e., Equation 1 and 2, and should be inserted using the {equation} environment, not as graphics
	% Please do not use footnotes in the text, additional information can be added to the Reference list.

	\section{Introduction}
	
	As one of the leading candidates for dark matter (DM), axions are lightweight pseudo-scalar particles. The Peccei-Quinn mechanism, which provides a solution to the strong charge-parity (CP) problem in Quantum Chromodynamics (QCD), involves a global U(1) symmetry that is spontaneously broken at a specific energy scale. The axion emerges as a pseudo-Nambu-Goldstone boson from this spontaneous symmetry breaking, and subsequently its potential arises from shift symmetry breaking.\textsuperscript{[\cite{Peccei:1977hh,Peccei:1977ur,Weinberg:1977ma,Wilczek:1977pj}]} Extensive experimental efforts have been devoted to the detection of the QCD axion and axion-like particles as dark matter. To date, the extremely weak interaction of axions make their direct detection exceptionally challenging.\textsuperscript{[\cite{Millar:2016cjp,Mitridate:2020kly,PhysRevD.101.096013,Budker:2013hfa,PhysRevLett.108.161803,Lee:2024toz,Kalia:2024eml,Sikivie:2009qn,PhysRevLett.58.1799,Gondolo:2008dd,Gramolin:2020ict,Reina-Valero:2024wqx,Diehl_2023,pandaxcollaboration2025searchmevscaleaxionlikeparticles,https://doi.org/10.1155/2017/6432354,PhysRevA.97.042506,Berlin_2021,Chao2023liu,Duan2022nuy,APEX:2024jxw,Chen:2020cbs,Gao:2022zxc,Rosenzweig_2024,Dome_2023,Zhu_2023,Jiang_2023}]} One of the primary strategies for axion detection is based on their interaction with electromagnetic fields, with existing experiments focusing mainly on the coupling between the axion field and electromagnetic fields,\textsuperscript{[\cite{Sikivie:1983ip}]} as well as the couplings to spin particles through precision measurements.\textsuperscript{[\cite{Wei_2025,xu_constraining_2024,qiu_axion_2023,PhysRevD.111.073010,PhysRevResearch.3.013205,Xu:2025lly,Rong:2017wzk,Jiang:2021dby,Chao:2024owf,Graham:2013gfa,Budker:2013hfa}]} This includes detecting the conversion of axions into microwave photons within a resonant cavity in a superconducting magnet, utilizing axion search telescopes to detect axions produced in the solar core, and using axion-photon oscillation experiments to detect axion signals.\textsuperscript{[\cite{Sikivie:1983ip,Irastorza:2018dyq,ParticleDataGroup:2018ovx,Lawson:2019brd,Sikivie:2020zpn,Rybka:2014xca}]} While these experiments have gradually improved the constraints on the coupling constant, a significant portion of the axion parameter space remains unexplored. This highlights the need for more innovative and advanced technological solutions to expand the search further and complement existing methods.
	
	In condensed matter systems, axions appear as quasi-particles in certain materials, such as topological insulators.\textsuperscript{[\cite{pellegrino2025localizedsurfaceplasmonsweyl,shyta2025axionelectrodynamicsweylsuperconductors,huang2025pressuredrivensuperconductivitytopologicalinsulator,filipini2025confinedelectromagneticwavesmedia}]} The corresponding axion fields in these systems give rise to various novel phenomena.\textsuperscript{[\cite{Nenno:2020ujq,Gramolin:2020ict,Li:2009tca,boyanovsky2025probingdynamicalaxionquasiparticles,shaposhnikov2024probingultrafastmagnetizationdynamics,boyanovsky2025syntheticcosmologicalaxionhybridization,prudêncio2022syntheticaxionresponsespacetime}]} While the axionic mode in condensed matter arises from a different nature, it offers the possibility for the system to respond to external axion-like perturbations, including dark matter. Recently, several studies have explored axion detection through resonance coupling with other quasi-particles, such as magnons, utilizing collective excitation modes and external magnetic fields that can modulate the detection.\textsuperscript{[\cite{PhysRevLett.123.121601,Catinari:2024ekq}]} Previous work has used multiferroic materials with spontaneous polarization and magnetization for detection, where the external dark matter axion field couples linearly with the ferromagnetic ordering in these materials, and the change in magnetization strength is measured.\textsuperscript{[\cite{Roising:2021lpv}]} However, multiferroic materials cannot directly detect axion dark matter due to their intrinsic material properties and coupling limitations.
	
\begin{figure}[t]
		\centering
		\includegraphics[width=1\textwidth]{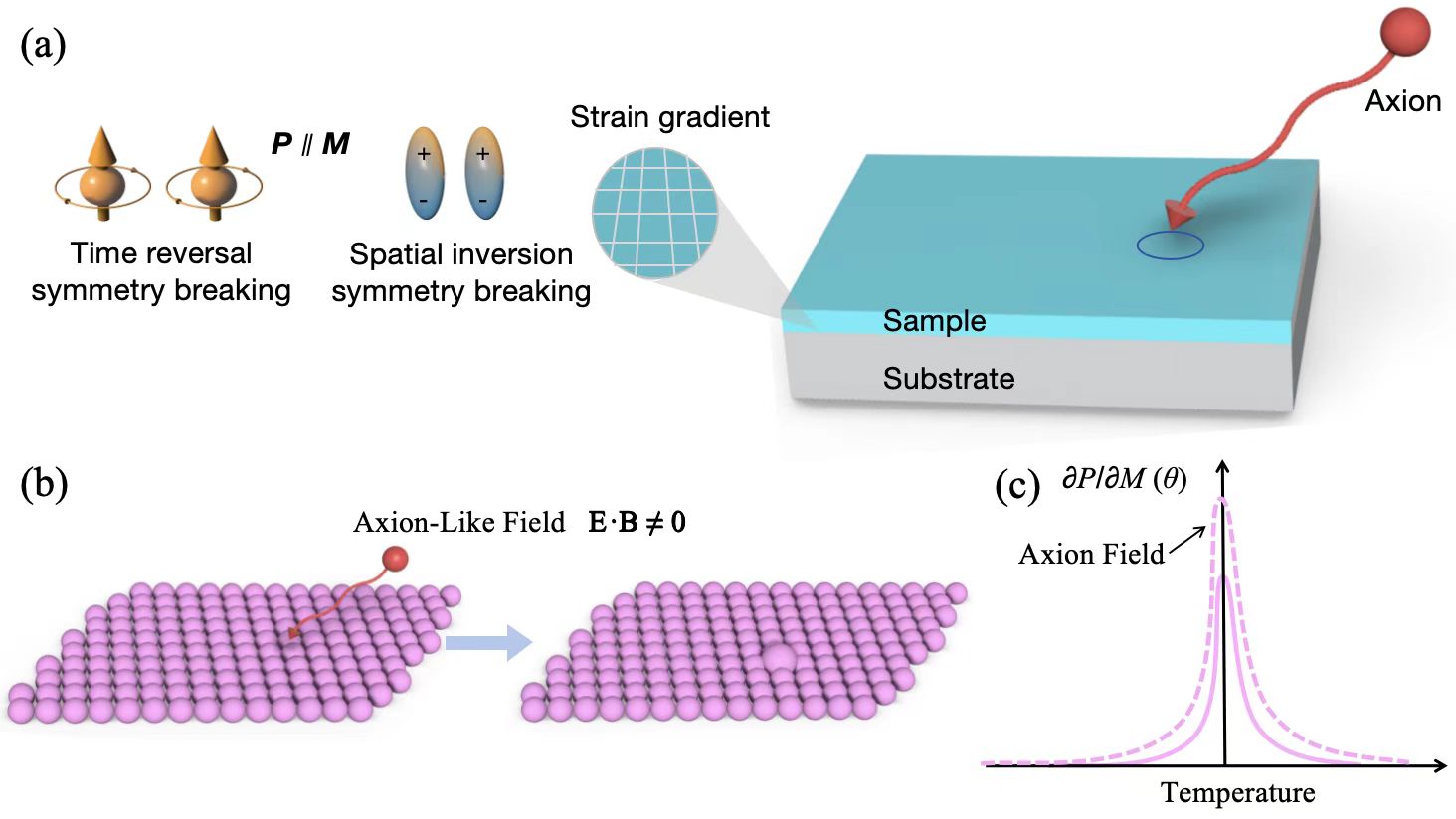}
		\caption{ (a) The magnetoelectric material exhibits parallel polarization (P) and magnetization (M), corresponding to the breaking of spatial inversion symmetry and time-reversal symmetry. (b) Enlarged view within the circle of (a), where the external axion field effectively couples with the internal axion-like field in the material. (c) Under axion dark matter excitation, the external axion field couples with the intrinsic P·M mode of the magnetoelectric material, modulating its magnetoelectric response.
		} 
		\label{fig:scheme}
	\end{figure}

	The magnetoelectric (ME) effect refers to a physical phenomenon where an external electric field can induce magnetization, or an external magnetic field can induce polarization in a material.\textsuperscript{[\cite{Rivera2009ASR,Fiebig2009CurrentTO,10.21468/SciPostPhys.6.4.046,article,HEHL20081141}]} This effect can be described by the equations $P_i=\alpha_{ij}H_j$ or $M_i=\alpha_{ji}E_j$ , where $P_i$ represents the polarization components, $M_i$ represents the magnetization components, $E_j$ and $H_j$ are the electric and magnetic field components, and $\alpha_{ij}$ is the magnetoelectric coupling tensor. Breaking time-reversal symmetry results in spin ordering, while breaking space-reversal symmetry leads to polar ordering.\textsuperscript{[\cite{TokuraKawasaki-4,2019arXiv190201532D}]} The simultaneous breaking of these symmetries allows for the mixing of polarization and magnetization, with the coupling between magnetism and spin order, giving rise to the magnetoelectric effect. When an axion field is present, the axion-photon interaction induces an oscillating electromagnetic field at the axion’s corresponding frequency. By applying a sufficiently strong static magnetic field $B_0$, axions can induce a weak oscillating electric field within this field. Magnetoelectric materials can convert this oscillating field into an observable polarization signal, thus enabling axion detection.\textsuperscript{[\cite{pellegrino2025localizedsurfaceplasmonsweyl,shyta2025axionelectrodynamicsweylsuperconductors,huang2025pressuredrivensuperconductivitytopologicalinsulator,filipini2025confinedelectromagneticwavesmedia,Ghosh_2024}]} Strong magnetic fields are commonly applied to generate a sufficiently large axion-photon conversion, as seen in the ADMX experiment.\textsuperscript{[\cite{ADMX:1998pbl}]} However, the application of a strong magnetic field may alter the domain structure of the magnetoelectric material or push it into a new magnetic ordered phase, potentially introducing indistinguishable spurious signals or weakening the intrinsic magnetoelectric coupling effect.\textsuperscript{[\cite{ueda_magnetic_2016}]} In this work, we propose a method to detect axions without applying strong magnetic fields, relying on materials with spontaneous symmetry breaking. Materials that allow magnetoelectric coupling can efficiently convert weak oscillating electromagnetic fields into changes in magnetization or polarization. Since the oscillating field induced by axions is extremely weak, establishing an axionic linear $\mathbf{P} || \mathbf{M}$ in the material —— that is, creating an axionic quasiparticle with parallel polarization and magnetization\textsuperscript{[\cite{Ghosh_2024,prudêncio2022syntheticaxionresponsespacetime}]}——allows for the hybridization\textsuperscript{[\cite{boyanovsky2025syntheticcosmologicalaxionhybridization}]} between the external axion field and the material’s internal axion-like field and this coupling enables the conversion of weak signals into observable quantities through a series of resonant or inductive measurements, as shown in Figure \ref{fig:scheme}. Magnetoelectric materials, in this case, provide a potential signal enhancement mechanism. Magnetoelectric materials can maintain symmetry breaking within a specific temperature or frequency range. They can be tuned through external fields or stress, making them ideal for scanning the unknown axion mass/frequency range.
	
In our recent work, we discovered a magnetoelectric phenomenon in $\text{Sr}_2\text{IrO}_4(\text{SIO})$ thin films through symmetry operations.\textsuperscript{ [\cite{PhysRevLett.133.156505}]} In this work, we propose this material as an excellent probe for axion dark matter. To demonstrate its potential, we employ two complementary experimental approaches: one based on SQUID magnetization measurements and another on direct magnetoelectric coupling measurement system (MECMS). These methods together provide a comprehensive framework for detecting axion-induced signals, as detailed in the following sections.
	\section{ Materials and Methods}
	Via the laser molecular beam epitaxy with in-situ reflection high-energy electron diffraction (RHEED), high-quality epitaxial SIO thin films were synthesized on $\text{Sr}\text{TiO}_3(\text{STO})$ (001) substrates with a thickness of 300 nm. 
	%High-quality $\text{Sr}_2\text{IrO}_4(\text{SIO})$ thin films were fabricated on single-crystal $\text{Sr}\text{TiO}_3(\text{STO})$ substrates using pulsed laser deposition (PLD) with in situ reflection high-energy electron diffraction (RHEED) monitoring.
	The films were grown at a substrate temperature of 830~°C with a heating rate of 20~°C/min under an oxygen pressure of 0.02 mbar. A KrF excimer laser with a wavelength of 248 nm, an energy density of 1 J/cm², and a pulse frequency of 2 Hz was used for the deposition. After deposition, the films were slowly cooled (5 °C/min) under an oxygen pressure of 100 mbar.
	
\begin{figure}[t]
		\centering
		\includegraphics[width=1\textwidth]{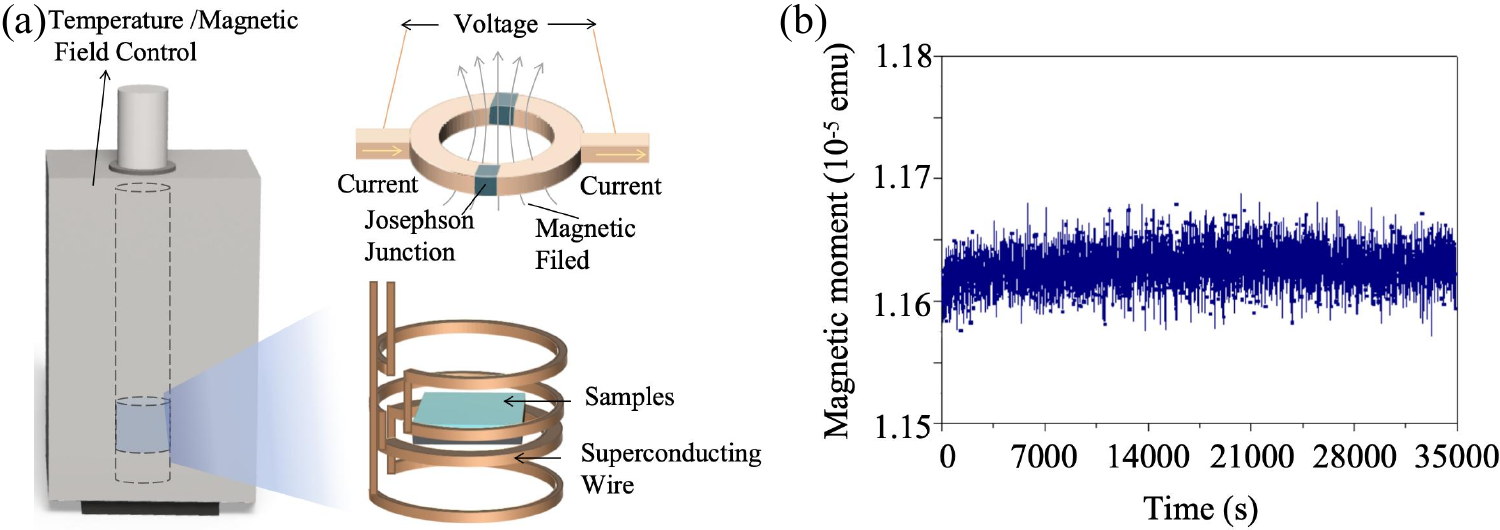}
		\caption{(a)  Schematic of the SQUID measurement system, when the sample moves within the superconducting detection coil, the magnetic flux through the coil changes, inducing an electromotive force that reflects the sample’s magnetic moment. The magnetic flux is directly coupled to the SQUID sensor, where the Josephson junctions convert the flux variation into a measurable voltage signal.  (b) Time-dependent magnetization curve at 20 K.
		} 
		\label{fig:SQUIDMM}
	\end{figure}

\begin{figure}[t]
		\centering
		\includegraphics[width=1\textwidth]{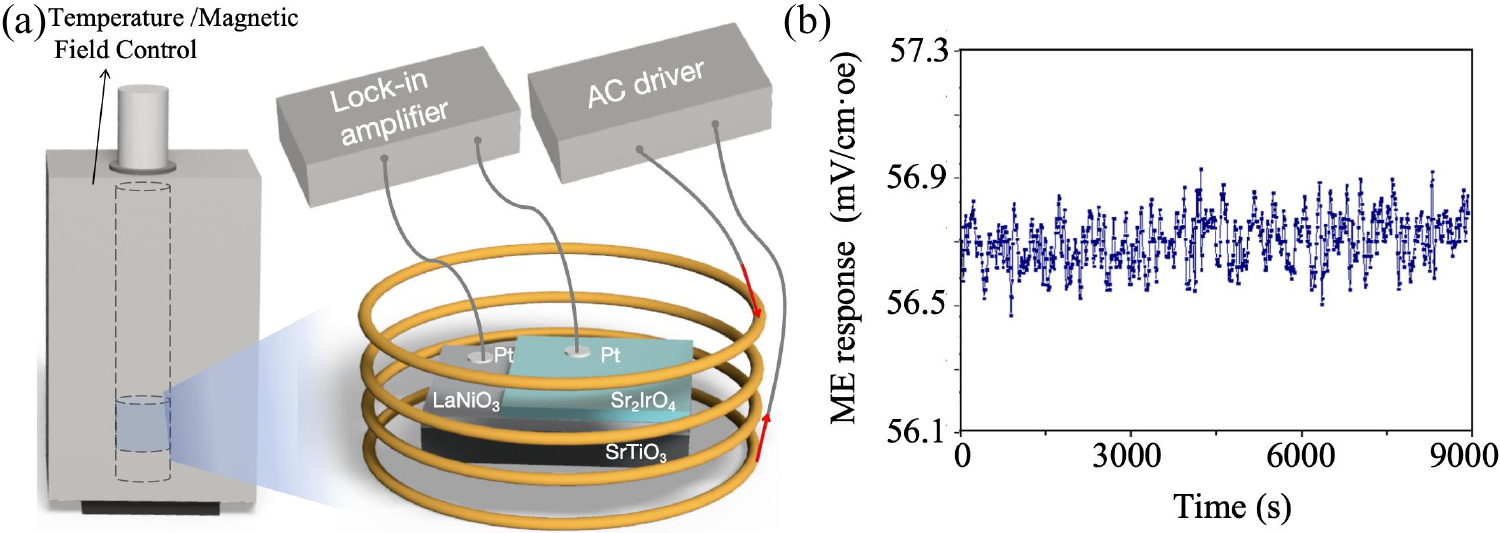}
		\caption{(a) Schematic of the magnetoelectric coupling measurement system. Electrodes are fabricated on the top and bottom surfaces of the $\text{Sr}_2\text{IrO}_4$ sample to collect electrical signals. The orange rings denote Helmholtz coils, which are driven by an AC power supply to generate an axial alternating magnetic field. Under this field, an alternating electrical signal is induced between the top and bottom electrodes of the sample and detected using a lock-in amplifier.   (b) Time-dependent magnetoelectric response curve at 20 K.
		} 
		\label{fig:MECMS}
	\end{figure}

	Magnetization information was obtained by using a superconducting quantum interference device (SQUID) magnetometer (Quantum Design) at 20 K. During the measurements, a DC magnetic field of 1000 Oe along the z-axis was applied as shown in the testing setup in Figure \ref{fig:SQUIDMM}a. The magnetoelectric response was measured in a physical property measurement system (PPMS, Quantum Design). The sample was cooled to 20 K to measure the magnetoelectric (ME) response. During the measurement process, an alternating magnetic field (\(\sim 2.5\) Oe) along the z-axis was generated by an alternating current source (Keithley 6221). The induced alternating voltage along the z-direction was measured by a lock-in amplifier (Stanford Research 830). The frequency of the ME measurement was set at 10 kHz. The schematic diagram of the magnetoelectric (ME) testing setup is shown in Figure \ref{fig:MECMS}a. 
	
	$\text{Sr}_2\text{IrO}_4$ is an iridate material characterized by strong spin-orbit coupling, exhibiting a rich spectrum of physical properties and generating a variety of novel electronic states.\textsuperscript{[\cite{2008PhRvL.101g6402K,PhysRevLett.101.226402}]} It adopts a single-layer perovskite structure with the RP phase and belongs to the tetragonal crystal system, space group I41/acd.\textsuperscript{[\cite{doi:10.1126/science.1167106}]} Each Ir ion is coordinated by six O ions, forming an $\text{IrO}_6$ octahedral structure. These octahedra are stacked in layers, with Sr ions intercalated between them. The $\text{IrO}_6$ octahedra rotate around the c-axis by 11°,\textsuperscript{[\cite{2008JPCM...20C5201K}]} which reduces the symmetry of the lattice structure. $\text{Sr}_2\text{IrO}_4$ exhibits characteristic physical properties strongly coupled with its lattice structure. The rotation of the $\text{IrO}_6$ octahedra plays a crucial role in the material's electronic and magnetic properties. $\text{Sr}_2\text{IrO}_4$ is antiferromagnetic, with a Néel temperature of approximately 240 K. The magnetization is coupled to the lattice, with the tilt and rotation angles of the octahedra being equal and locked.\textsuperscript{[\cite{2013PhRvB..87n0406Y}]}  Due to the tilting of the magnetic moment, $\text{Sr}_2\text{IrO}_4$ exhibits a net magnetic moment within each $\text{IrO}_2$ plane, alternating between layers. Along the c-axis, the material displays an antiferromagnetic arrangement, compensating interlayer magnetic moments, resulting in no macroscopic magnetization. However, when an external magnetic field exceeds a critical threshold, the material’s critical field, $\text{H}_{\text{fiop}}$, is relatively small (1000 Oe), which induces a phase transition and results in a net magnetic moment in the $\text{IrO}_2$ layers, exhibiting weak ferromagnetism.\textsuperscript{[\cite{2019PhRvB..99h5125P,doi:10.1126/science.1167106}]}  The Ir-O-Ir bond angle can be modulated by external factors such as magnetic fields, pressure, electric fields, and epitaxial strain.\textsuperscript{[\cite{Jackeli:2009qje,PhysRevLett.114.096404}]} Additionally, constructing superlattices can also lead to a tunable ground state.\textsuperscript{[\cite{LiuSong-3}]} On a related note, dipole-dipole repulsion imposes constraints on dipole density. Elevated dipole densities induce either depolarization or, subject to the availability of free charge carriers, a transition from insulating to metallic behavior.\textsuperscript{[\cite{PhysRev.33.954}]} Notwithstanding dipole density implications raised, the material demonstrates persistent insulating characteristics without evidence of metallic transition.
	
	Our recent work achieved a magnetoelectric effect in strain gradient $\text{Sr}_2\text{IrO}_4$ thin films. The space inversion and time reversal symmetries were manipulated by artificially designing graded strain in SIO films, resulting in emergent polar ordering and unconventional magnetism along z axis. These two orderings combine by a non-equivalent pd hybridization to drive a flexomagnetoelectric effect.  Furthermore, the material exhibits excellent insulation retention under cryogenic conditions. This ME response exhibits a maximum output at \(\sim\) 20 K, where a polar phase transition occurs. \textsuperscript{[\cite{PhysRevLett.133.156505}]}  In addition, this symmetry engineering results in a parallel polarization and magnetism in SIO, contributing to axion detection. The unique properties of SIO allow us to explore axion interactions through two distinct experimental setups: (1) a SQUID-based magnetization measurement and (2) a direct magnetoelectric coupling measurement. These approaches are designed to capture different aspects of the axion-material interaction, as elaborated in the following sections.
	
	The choice of $\text{Sr}_2\text{IrO}_4$ for axion detection is particularly advantageous for several reasons. First, its strong spin-orbit coupling and flexomagnetoelectric response outperform other magnetoelectrics in coupling efficiency. Second, the strain-gradient design in thin films breaks inversion symmetry in a controllable manner, enabling a linear  $\mathbf{P} || \mathbf{M}$ coupling that directly interfaces with the axion field - a feature not achievable in bulk materials or conventional magnetoelectrics. While traditional cavity-based methods rely on large volumes to enhance signals, our approach leverages these intrinsic material properties for signal amplification. As shown below, the specialized thin-film nature of our material already provides significant advantages in terms of coupling strength and tunability.
	\section{Detecting Axion by Measuring Magnetization}
	The axion  and the axion coupling to the electromagnetic field are described by:
	\begin{align}
		\mathcal{L}_a &= \frac{1}{2}(\partial_\mu a)(\partial^\mu a) - \frac{1}{2}m_a^2 a^2,\\
		\mathcal{L}_{a\gamma\gamma} &= -\frac{\theta_{a\gamma\gamma}}{4} F_{\mu\nu} \tilde{F}^{\mu\nu} = \theta_{a\gamma\gamma} \mathbf{E} \cdot \mathbf{B},\\
		\theta_{a\gamma\gamma} &\equiv a g_{a\gamma\gamma},\label{eq:theta}
	\end{align}
	where $\theta_{a\gamma\gamma}$ can be regarded as a mixing between the electric field and magnetic field, and it is related to the axion-photon coupling $g_{a\gamma\gamma}$ in the presence of a background axion field. \(F^{\mu\nu}\) is electromagnetic field tensor, \(\tilde{F}^{\mu\nu}=\frac{1}{2}\epsilon^{\mu\nu\alpha\beta}F_{\alpha\beta}\) is electromagnetic dual tensor, $\epsilon^{\mu\nu\alpha\beta}$ is Levi-Civita symbol and \(\epsilon^{0123}=1\). In the zero-velocity limit, the axion field \(a\) is described by an amplitude $a_0$ and frequency $\omega_a=m_a$, which leads to the form: $a(t)=a_0e^{-i\omega_at}$.
	The physical axion field corresponds to the real part of this expression. The amplitude is determined by the local dark matter density, given by:\textsuperscript{[\cite{Millar:2016cjp}]}
	\begin{align}
		\rho_{\text{DM}}=\frac{1}{2}m_a^2|a_0|^2,\label{eq:rho}
	\end{align}
	where $\rho_{\text{DM}}$ is the local DM density and $\rho_{\text{DM}}\approx 0.4~ \text{GeV} \text{cm}^{-3}$. In magnetoelectric systems, the electric field and magnetic field within the material are aligned in parallel due to the breaking of time symmetry and inversion symmetry. Therefore, the axion can couple linearly with the electric and magnetic fields within the material. In Ginzburg-Landau theory, the coupling between the axion and the
	$\mathbf{P}\cdot\mathbf{M}$ term in the ME material needs to be considered,  so the term $\chi\theta\mathbf{P}\cdot\mathbf{M}$ should be included in the free energy density.\textsuperscript{[\cite{Roising:2021lpv}]} The electromagnetic coupling of the material itself is described by $\alpha_{ij}$. According to the Ginzburg-Landau theory, it can be concluded that:\textsuperscript{[\cite{Roising:2021lpv}]}
	\begin{align}
		\mathcal{F}[\mathbf{M}, \mathbf{P}] &= \mu_0 \mathcal{F}_M[\mathbf{M}] + \frac{1}{\varepsilon_0}  \mathcal{F}_P[\mathbf{P}] - \sqrt{\frac{\mu_0}{\varepsilon_0}}(\alpha_{ij} M^{i} P^{j} + \chi \theta  \mathbf{P} \cdot \mathbf{M} ) , \label{eq:F}\\
		\mathcal{F}_M[\mathbf{M}] &= \alpha_M \lvert \partial_t \mathbf{M} \rvert^2 - \beta_M \lvert \mathbf{\nabla} \cdot \mathbf{M} \rvert^2 -  \gamma_M(T-T_{C_M}) \lvert \mathbf{M} \rvert^2 - \lambda_M \lvert \mathbf{M} \rvert^4 + \dots, \\
		\mathcal{F}_P[\mathbf{P}] &= \alpha_P \lvert \partial_t \mathbf{P} \rvert^2 - \beta_P \lvert \mathbf{\nabla} \cdot \mathbf{P} \rvert^2 - \gamma_P(T-T_{C_P}) \lvert \mathbf{P} \rvert^2 - \lambda_P \lvert \mathbf{P} \rvert^4 + \dots,
	\end{align}
	where $\varepsilon_0$ ($\mu_0$) is the vacuum permittivity (permeability), and $\alpha$'s, $\beta$'s, $\gamma$'s, and $\lambda$'s are phenomenological coefficients, and $\chi$ is a material dependent susceptibility. The transition temperature ${T}_{C_M}$ (${T}_{C_P}$) denotes the magnetic electric phase transition temperature, $\alpha_{ij}$ is the material dependent ME tensor. For simplicity, we only consider the response of $\mathbf{P}||\mathbf{M}$ in the material with $\alpha_{ij}=\alpha\delta_{ij}$. The axion term $\theta(t)=\theta_0 e^{-i\omega_at}$ can be regarded as a weak driving force that varies with time.
	
   The classical equations of motion for $\bf{M}$ and $\bf{P}$:\textsuperscript{[\cite{Roising:2021lpv}]}
	\begin{align}
	\alpha_M \partial_t^2 \mathbf{M} - \beta_M \nabla^2 \mathbf{M} + \gamma_M (T-T_{C_M}) \mathbf{M} &  \nonumber \\
	+ \frac{c}{2}(\underline{\alpha} + \chi \theta) (\mathbf{P}+\mathbf{P}_0) + 2\lambda_M \lvert \mathbf{M} \rvert^2 \mathbf{M} &= 0, \label{eq:ELeqM} \\
	\alpha_P \partial_t^2 \mathbf{P} - \beta_P \nabla^2 \mathbf{P} + \frac12 m_P^2\mathbf{P} + \frac{1}{2c}(\underline{\alpha} + \chi \theta) \mathbf{M} &= 0, \label{eq:ELeqP}
	\end{align}
	where $\underline{\alpha}$ is the ME tensor with elements $\alpha_{ij}$. We consider spatially homogeneous solutions of the longitudinal modes of Equation \ref{eq:ELeqM} and \ref{eq:ELeqP}:
	$\mathbf{P}(t;x)=P(t)\hat{e}, \mathbf{M}(t;x)=M(t)\hat{e}$. $M_0$ and $P_0$ denote the values of  static magnetization and polarization in the absence of the axion field. Then we linearize the  equations by $M=M_0+\delta M, P=P_0+\delta P$,  neglect the second-order terms in the equation and also manually add phenomenological damping terms $\eta_M$ and $\eta_P$. We set $\delta M(t)=\delta M\ \text{exp}(-i\omega_a t)$, $\delta P(t)=\delta P\ \text{exp}(-i\omega_a t)$ and we can derive that:
	\begin{align}
		\delta M&=\frac{\theta_0 \chi  \left(\alpha  M_0-2 c  P_0 \left(i\alpha_P \eta_P \omega_a-\alpha_P\omega_a^2+m_P^2\right)\right)}
		{-\alpha ^2+4  \left(i\alpha_M \eta_M \omega_a-\alpha_M\omega_a^2+m_M^2\right) \left(i\alpha_P \eta_P \omega_a-\alpha_P\omega_a^2+m_P^2\right)},\\
		\delta P&=\frac{\theta_0 \chi  \left(\alpha c P_0-2  M_0 \left(i\alpha_M  \eta_M \omega_a-\alpha_M \omega_a^2+m_M^2\right)\right)}
		{-\alpha ^2+4\left(i \alpha_M\eta_M \omega_a-\alpha_M\omega_a^2+m_M^2\right) \left(i\alpha_P \eta_P \omega_a-\alpha_P\omega_a^2+m_P^2\right)},
	\end{align}
	where \(m_M\) and \(m_P\) are artificially defined dimensionless effective mass parameters, introduced to characterize the frequency response of the magnetization and polarization order parameters near their equilibrium states and defined by $m_M^2 \equiv\gamma_M(T-T_{C_M}) + 6 \lambda_M M_0^2$, $m_P^2 \equiv \gamma_P(T-T_{C_P}) + 6 \lambda_P P_0^2$.\textsuperscript{[\cite{Roising:2021lpv}]} $\delta M$ and $\delta P$ have the same form of solution with tranformation: $M_0\leftrightarrow cP_0, \alpha_M\leftrightarrow \alpha_P, \eta_M\leftrightarrow \eta_P, m_M\leftrightarrow m_P$.
	
\begin{figure}[t]
		\centering
		\includegraphics[width=0.8\textwidth]{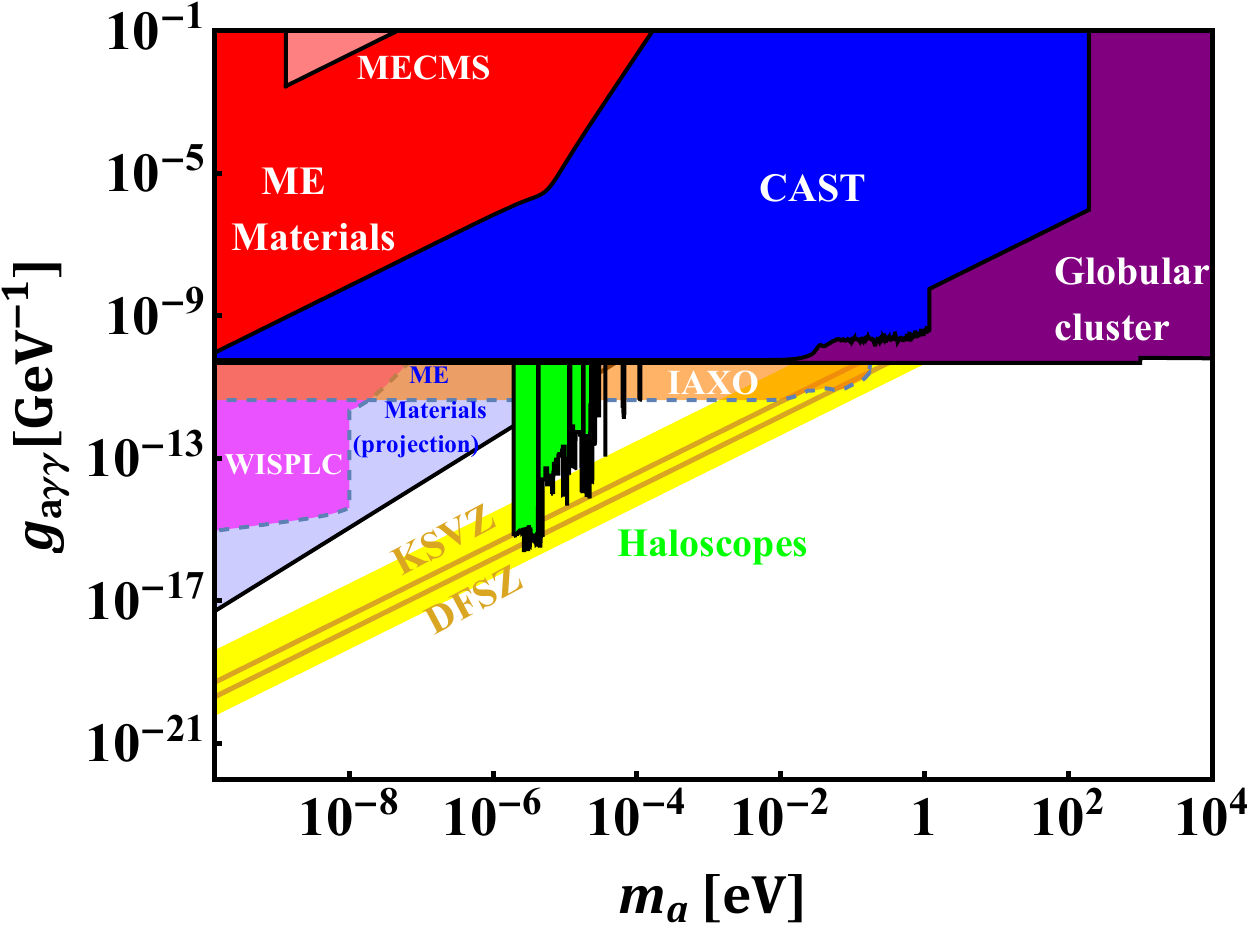}
		\caption{Constraints on the coupling factor $g_{a\gamma\gamma }$. 
			The red and pink exclusions are obtained in our experiments, which utilize an external magnetic field of approximately 1000 Oe. The size of the test sample is $300~\text{nm}\times2.5~\text{mm}\times2.5~\text{mm}$. The SQUID measurement system (ME Materials) obtains the red exclusion, and the magnetoelectric coupling measurement system (MECMS) obtains the pink exclusion. For the red exclusion, the integration time of $M$ is 10 seconds, while for the pink exclusion, the integration time of $\theta$ is 1 second. The projection of Magnetoelectric Materials is based on an SNR=3 criterion. The size of the magnetoelectric material sample for the projection is $0.1\times0.1\times0.01~\text{m}^3$. The integration time of the projection is one year. We consider the 95$\%$ C.L. exclusion limit.
			The bound of the Globular cluster comes from Reference \cite{Ayala:2014pea}. The bounds of CAST come from Reference \cite{CAST:2024eil}. The projection of WISPLC is based on Reference \cite{Zhang:2021bpa}. The projection of IAXO comes from Reference \cite{IAXO:2019mpb}. The yellow lines represent Kim-Shifman-Vainshtein-Zakharov (KSVZ) and Dine-Fischler-Srednicki-Zhitnitsky (DFSZ)  axion models, and the yellow region represents the hadronic band. } 
		\label{fig:garr}
	\end{figure}

	Here, we write down the response of the magnetization coupled with the polarization as a function of the frequency of the axion:
	
	\begin{small}
		\begin{align}
           &\left|\frac{\delta M}{\theta_0}   \right|=\notag \\ 
			&\frac{\chi \sqrt{4c^2P_0^2\alpha_P^2\eta_P^2\omega_a^2+[M_0\alpha+2cP_0(\alpha_P\omega_a^2-m_P^2)]^2}}
			{\sqrt{16\omega_a^2[\alpha_M\alpha_P\omega_a^2(\eta_M+\eta_P)-\alpha_P\eta_Pm_M^2-\alpha_M\eta_Mm_P^2]^2
					+[\alpha^2+4\alpha_M\alpha_P\eta_M\eta_P\omega_a^2-4(\alpha_M\omega_a^2-m_M^2)(\alpha_P\omega_a^2-m_P^2)]^2}}.\label{eq:deltaM1}
		\end{align}
	\end{small}
	%The solution of $|\delta P/\theta_0|$ has the same form by the tranformation above. 
	In our experiment, $\alpha_M=1~\text{meV}^{-2}$ and $\alpha_P = 10^{-2}~\text{meV}^{-2}$ are phenomenological coefficients. $\eta_M=5~\mu\text{eV}$ and $\eta_P=5~\mu\text{eV}$ are damping coefficients, $\chi=\frac{M}{H}\simeq0.03$ is a material dependent susceptibility.  $H\simeq1000~\text{Oe}$ is magnetic field intensity.\textsuperscript{[\cite{Roising:2021lpv}]}
	$P_0=P-\delta P=0.6~\mu\text{C}/\text{cm}^2$ and $M_0=M-\delta M=4.54\times10^{-6}~\text{emu}/\text{Vol}$ are the values of the static polarization and magnetization in the absence of the axion field. The measurement of $M$ is shown in Figure \ref{fig:SQUIDMM}b, therefore $\delta M=1.30419\times10^{-8}~\text{emu}/\text{Vol}$. Vol$=300 ~\text{nm}\times 2.5~\text{mm}\times 2.5~\text{mm}$ is the size of magnetoelectric material. c is the speed of light in a vacuum, and nonrelativistically $\omega_a\approx m_a$ is the axion's mass. $m_P=4.5\times10^{-2}$  is the effective mass at equilibrium. $\gamma_M=10^{-5}~\text{K}^{-1}$ and $\lambda_M=9.25~\text{kOe}^{-2}$ are phenomenological coefficients. The transition temperature $T_{C_M}=20~\text{K}$ denotes the magnetic phase transition temperature. In our experiments, we measured the magnetization $M$ of the material, and the standard error formula was used to determine $\delta M$. From the Equation \ref{eq:deltaM1}, we can infer \(\theta_0\), ultimately constraining \(g_{a\gamma\gamma}\) through Equation \ref{eq:theta} and \ref{eq:rho}. This allows us to obtain the red exclusion line of \(g_{a\gamma\gamma}\) in Figure \ref{fig:garr}. Since the numerator and denominator on the right side of Equation \ref{eq:deltaM1} have the same power of $\omega_a$, and given that Equation \ref{eq:theta} ($g_{a\gamma\gamma}\sim a^{-1}$) and Equation \ref{eq:rho} ($a\sim m_a^{-1}$), the exclusion line has a positive slope ($g_{a\gamma\gamma}\sim m_a \theta_{a\gamma\gamma}$). Theoretically, this implies a higher constraint at lower frequencies.
	
	Microscopically, the effective $\mathbf{P}\cdot\mathbf{M}$ coupling can also be derived from the axion coupling to electrons, as shown in the following equation:\textsuperscript{[\cite{Roising:2021lpv}]}
	\begin{align}
		\chi \theta_{ae} &\equiv g_{ae}  \frac{a_0 m_a \hbar}{ m_e^2 c^2 g_s \alpha_E e^2} \sqrt{ \frac{\varepsilon_0}{\mu_0} } \sim 7.7 \times 10^{-11} g_{ae}, \label{eq:EffectiveCoupling} \\
		\delta E_{ae}^{\mathrm{tot}} &\approx L_{\mathrm{domain}}^3  \chi \theta_{ae} \sqrt{\frac{\mu_0}{\varepsilon_0}}\mathbf{P}\cdot\mathbf{M},\label{eq:Eae}
	\end{align}
	where $m_e$ is the mass of electron, $g_s\simeq2$ is the spin g-factor of electrons, and $\alpha_E \simeq1/137$ is the fine structure constant, $\mu_0$ and $\epsilon_0$ are  the permeability and permittivity of free space respectively, \(\delta E_{ae}^{\mathrm{tot}}\) is the energy change caused by axion, \(c\) is the speed of light in a vacuum, $L_{\text{domain}}$ is the domain size. By comparing Equation \ref{eq:F} with Equation \ref{eq:Eae}, we consider the axion field as a small perturbation to the material near its equilibrium state. Using Equation \ref{eq:deltaM1} and Equation \ref{eq:EffectiveCoupling},
	we can derive the red exclusion area for $g_{ae}$, as shown in Figure \ref{fig:gae}.
	% Since the SQUID makes the sample cut the magnetic inductance line at a vibration frequency of 40 Hz, we have a energy threshold $\Lambda\simeq8.26\times10^{-14}$~eV.
	
	We assume that the de Broglie length of the axion is larger than the material size and neglect the spatial dependence of the axion-like particle (ALP) background inside the material.\textsuperscript{[\cite{chigusa_axionhidden-photon_2021}]} Thus the mass of axion must satisfy  $m_a<\frac{2\pi\hbar v_a}{l}$, where $l$ is the thickness of ME material and $v_a=10^{-3}$ is the velocity of local DM. While the SQUID-based magnetization measurement provides a sensitive probe of axion-electron coupling, it is complemented by the direct magnetoelectric coupling measurement, which offers an alternative pathway to constrain axion-photon interactions, as discussed in the next section.

	\section{The Magnetoelectric Coupling}
    Building on the magnetization measurements discussed earlier, we now turn to the direct characterization of magnetoelectric coupling in SIO. Since the material exhibits intrinsic magnetoelectric coupling, we can use the parameter \(\theta_{\text{ME}}\)
	to describe this property. This approach not only provides an independent validation of the SQUID results but also extends the detection sensitivity to different axion coupling regimes. The intrinsic magnetoelectric coupling of the material is defined by \(\theta_{\text{ME}}\), and the change in magnetoelectric coupling induced by the dynamical axion field is denoted by \(\theta\). Therefore, the total magnetoelectric coupling of the material is given by:
	\begin{align}
		\theta_{tot}=\theta_{\text{ME}}+\theta.\label{eq:theta1}
	\end{align}
	where the latter $\theta$ is due to the influence of axion dark matter. In condensed matter systems, $\theta_{tot}$ can be viewed as a pseudoscalar field associated with axion quasiparticles.\textsuperscript{[\cite{Sekine:2020ixs}]}
	Within the axionic linear ME response region, the linear magnetoelectric coupling coefficient is generically described by:
	%In natural units, permeability of vacuum $\mu _0=1$. Magnetic susceptibility $\chi_m\sim10^{-3}\ll 1$, which means relative permeability $\mu_r=1+\chi_m\simeq 1$.
	\begin{align}
		\alpha_{ij}= \left.\frac{\partial P_i}{\partial H_j}\right| _{\mathbf{E}=0}.
	\end{align}
	The magnetoelectric coupling 
	%in multiferroics
	has the relation:\textsuperscript{[\cite{Roising:2021lpv,Sekine:2020ixs,eerenstein_multiferroic_2006,PhysRevLett.102.146805,PhysRevLett.49.405,PhysRevB.78.195424,PhysRevLett.58.1799,hehl_magnetoelectric_2009,Schütte-Engel_2021,chigusa_axionhidden-photon_2021,qiu_observation_2025,PhysRevLett.123.121601}]}
	\begin{align}
		\theta_{\text{ME}}&=\frac{4\pi^2\hbar c\alpha}{e^2}.
	\end{align}
	%Where $\theta$ is the dimensionless “axion angle".
	%%%[On the matter of topological insulators as magnetoelectric]
	%We call $\theta$ an axion field in analogy with its interpretation as a fundamental gauge potential in high-energy physics. Its quantized excitations are consequently called axions. In the condensed-matter context, $\theta$ can also be interpreted as a contribution to the magnetoelectric polarization from extended orbitals.\\
	%The field 
	%$\theta$ is often referred to as an axion field, drawing a parallel with its role as an elementary gauge potential in high-energy particle physics. The discrete energy states associated with this field are correspondingly termed axions. Within the framework of condensed matter physics, 
	%$\theta$ can alternatively be understood as a component of magnetoelectric polarization arising from delocalized electronic orbitals.\\
	Experimentally, we measured the data of $\frac{\partial P}{\partial H}|_{\mathbf{E}=0}$,  as shown in Figure \ref{fig:MECMS}b. The average value corresponds to $\theta_{\text{ME}}$, and the standard error corresponds to $\theta$. Thus, we obtained theta and the pink limits of \(g_{ae}\), as shown in Figure \ref{fig:gae}. In the same way, we obtained the pink exclusion line in Figure \ref{fig:garr}. 
	%Since the frequency of ME measurement was set at 10~kHz, there is an energy threshold $\Lambda_{\text{ME}}\simeq4.13\times10^{-11}$~eV.

	\section{Discussion and Future Projection}
\begin{figure}[t]
		\centering
		\includegraphics[width=0.8\textwidth]{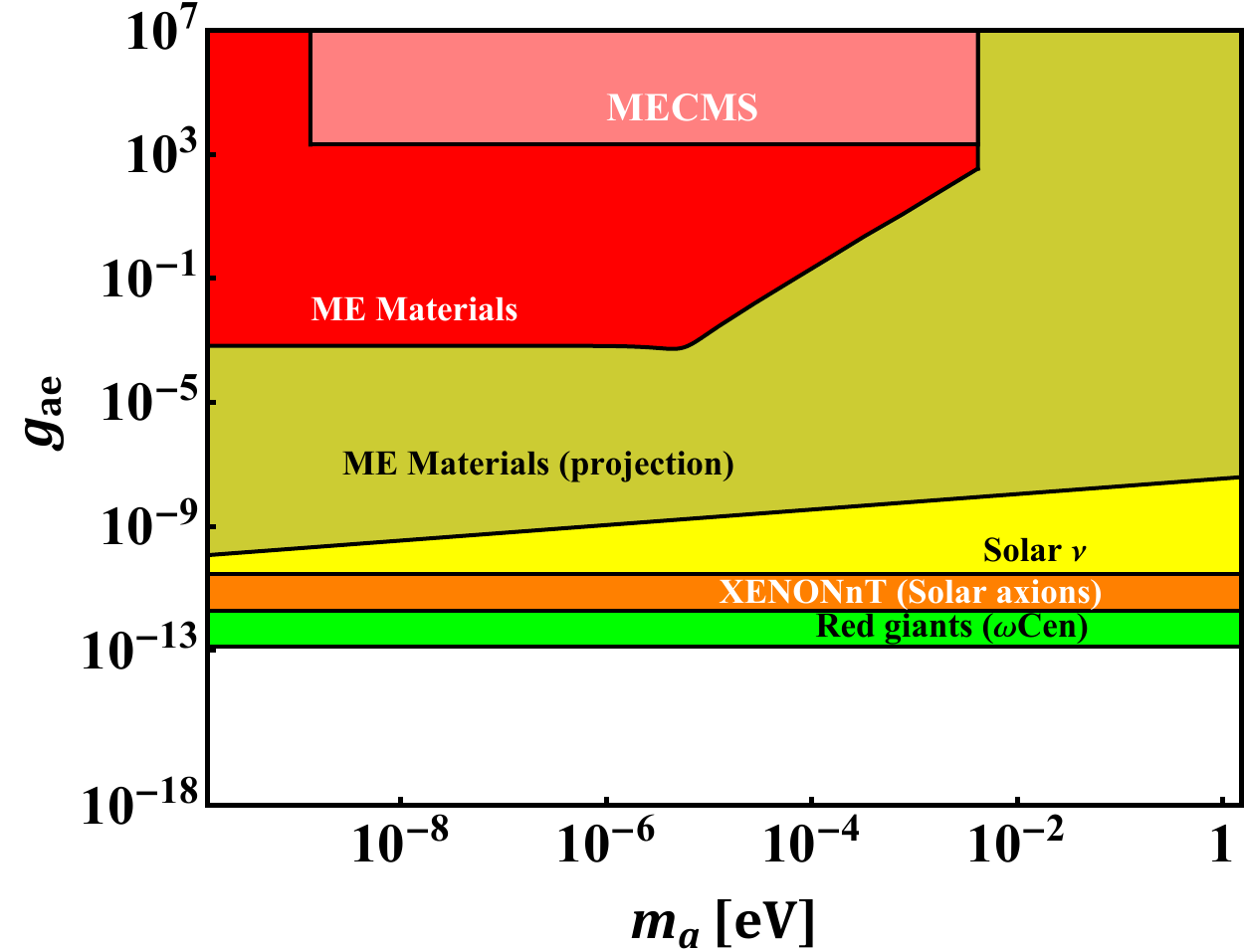}
		\caption{Constraints on the coupling factor $g_{ae }$. 
			The red and pink exclusion is obtained in our experiments, which utilize an external magnetic field of approximately 1000 Oe. The SQUID measurement system obtains the red exclusion, and the magnetoelectric coupling measurement system (MECMS) obtains the pink exclusion. The size of the test sample is $300~\text{nm}\times2.5~\text{mm}\times2.5~\text{mm}$. For the red exclusion, the integration time of $M$ is 10 seconds, while for the pink exclusion, the integration time of $\theta$ is 1 second.
			The projection of Magnetoelectric Materials is made assuming SNR=3. The size of the magnetoelectric material sample for the projection is $0.1\times0.1\times0.01~\text{m}^3$. The integration time of the projection is one year.  We consider the 95$\%$ C.L. exclusion limit. The bounds of XENONnT (Solar axions) and XENONnT (ALP DM) come from Reference \cite{XENON:2022ltv}. The Red giant branch bound is taken from Reference \cite{Capozzi:2020cbu}. The constraint of Solar neutrinos comes from Reference \cite{Gondolo:2008dd}. 
		} 
		\label{fig:gae}
	\end{figure}

	To sum up, the exclusion line can be determined through two distinct yet complementary approaches: the SQUID-based magnetization measurement and the direct magnetoelectric coupling measurement. The former excels in probing axion couplings, while the latter provides a novel and simpler way for axion couplings detection. Together, they form a robust framework for axion detection, as illustrated in Figures 4 and 5.The first method involves measuring $M$ to obtain $\delta M$, subsequently deriving $\theta$ using Equation \ref{eq:deltaM1}, and finally calculating $g_{ae}$ through Equation \ref{eq:EffectiveCoupling} or $g_{a\gamma\gamma}$ through Equation \ref{eq:theta}. Alternatively, the second method entails measuring $\left.\frac{\partial P}{\partial H}\right|_{\mathbf{E}=0}$ to acquire the axionic linear magnetoelectric coupling coefficient $\alpha$, followed by the determination of $\theta_{\text{ME}}$ and $\theta$, ultimately yielding $g_{ae}$ or $g_{a\gamma\gamma}$. As shown in Figure \ref{fig:gae}, although the first method is more restrictive, the second Original method is computationally and experimentally simpler. We can tighten the limit by superimposing the number of layers of ME material films and increasing the measurement time.
	%These two methods have a complementary relationship in the detection frequency band. The first detection method improves detection accuracy at specific frequency bands through resonance, while the second detection method is a wideband detection method. 
	
	For the first measurement method, based on existing literature, we have further considered the coupling between \( P \) and \( M \) and probed axions using novel materials at lower phase transition temperatures. The second method represents the first experimental attempt to constrain the axion coupling coefficient by directly measuring the magnetoelectric coefficient of magnetoelectric materials. Although the constraints are relatively weak, they provide valuable insights and inspiration for developing new axion detection experiments.
	
	One key advantage of using magnetoelectric materials for axion detection is that the axion's $\theta$ parameter couples to the material's intrinsic $\theta_{\text{ME}}$, regardless of the axion's specific theoretical models or interaction types. The constraints on the axion-electron coupling from magnetoelectric materials are weaker than those on the axion-photon-photon coupling. This is due to the suppression term $\frac{m_a}{m_e^2}$ in Equation \ref{eq:EffectiveCoupling}, which reduces the axion-electron coupling, whereas no such suppression exists for the axion-photon-photon coupling.
	The experiment sensitivity  for the first method to detect axion is characterized by the signal-to-noise ratio (SNR):
	\begin{equation}
		\begin{aligned}
			\text{SNR} &\sim\, \frac{V}{\gamma T} \frac{|P_0|^2}{\Delta\omega_a\alpha_M}(\chi\theta)^2\sqrt{\frac{\Delta\omega_a}{\Delta\omega_{\mathrm{meas}}}},\\
			%		&= 6.16177\times10^{18} \left( \frac{V}{300 \mathrm{nm}\times2.5\text{mm}^2} \right)\left( \frac{10^{-5}\text{eV}}{\gamma} \right)\left( \frac{20\mathrm{K}}{T} \right)\sqrt{\frac{1}{\Delta\omega_a\Delta\omega_{\mathrm{meas}}}}
		\end{aligned}
		\label{eq:SNBenjo}
	\end{equation}
	where $\gamma \sim 10^{-6}$~eV is the width of the magnetic resonance.\textsuperscript{[\cite{Balatsky:2023mdm}]}
	$\Delta\omega_a \sim \frac 12\omega_a v_a^2 $ is the width of the axion signal. $v_a\simeq10^{-3}$ is the velocity of local dark matter axions.
	%, expected to be determined by Doppler broadening of the Galactic axion background .
	$\Delta\omega_{\mathrm{meas}}\sim t_{\mathrm{meas}}^{-1}$ the measurement bandwidth, which is determined by the measurement time $t_{\mathrm{meas}}$: this estimate assumes $t_{\mathrm{meas}}= 1~\text{year}$.
	V is the size of the projected Magnetoelectric Material sample, which is $0.1\times0.1\times0.01$~m$^3$.
	%这就说明squid也得足够大从而能够去测量这么大的样品
	Here, we set SNR = 3 to get the expected projection of $g_{ae}$ and $ g_{a\gamma\gamma}$ of axion in ME materials in the case of one year of exposure, as illustrated in Figure \ref{fig:garr} and Figure \ref{fig:gae}. The projected sensitivity becomes competitive with state-of-the-art methods while eliminating the need for strong external magnetic fields - a key innovation of our approach.
	
	%Large V means that the SQUID also needs to be large enough to measure such a large sample.\\
	%The current experimental conditions do not allow the magnetoelectric material to be increased to 100 m$^3$. Today's lab equipment allows for a maximum of 20 layers of magnetoelectric film to be stacked into SQUID for testing. At the same time, the effect of increasing the volume of the test material on the change in magnetization M is not clear, so this projection is only a rough estimate of the order of magnitude, which is used to compare the results of existing experiments by orders of magnitude.
	For the time-domain signal of axion, its spectral width \( \Delta \omega_a \) determines the coherence time $t_{\text{coh}} \sim \frac{1}{\Delta \omega_a}$.
	\\If the measurement time exceeds the coherence time, repeated measurements can be treated as independent experiments. According to statistical principles, the precision (SNR or sensitivity) improves with the square root of the number of independent measurements. Since the number of independent measurements is \( t_{\mathrm{meas}}/t_{\text{coh}} \), the precision scales as:
	\[
	\text{SNR} \propto \left( \frac{t_{\mathrm{meas}}}{t_{\text{coh}}} \right)^{1/2}.
	\]
	The signal phase remains coherent for measurement times shorter than the coherence time. Here, precision is typically limited by quantum mechanical or wave-like properties. For certain systems (e.g., interferometers or resonant detectors), the precision may scale as \( t^{1/4} \) until it saturates at \( t \approx t_{\text{coh}} \). This behavior may depend on the type of noise (thermal or shot noise).
	For dark matter fields, if their interactions do not involve spatial derivatives, the coherence time is extremely long, typically:
	\[
	t_{\text{coh}} \sim 10^6 \cdot \left( \frac{2\pi}{\omega_a} \right),
	\]
	where \( \omega_a \) is the angular frequency of the dark matter field.
	If the interaction includes spatial derivatives, the coherence time is significantly reduced due to spatial inhomogeneities, potentially much shorter than the derivative-free case.
	
	To ensure that the measurement time $t_{\mathrm{meas}}$ exceeds the axion coherence time $t_{\text{coh}}$, the measurement bandwidth $\Delta\omega_{\mathrm{meas}}$ must be smaller than the bandwidth of the axion signal $ \Delta\omega_{a}$. Therefore, the axion mass must satisfy  $\omega_a>\Lambda_1\simeq1.314\times10^{-10}$~eV for the first approach to detect axion and  $\omega_a>\Lambda_{2}\simeq1.314\times10^{-9}$~eV  for the second one, where $\Lambda_{1(2)}=\frac{2}{t_{1(2)} v_a^2}$ is an energy threshold with $t_{1}=10$~s for SQUID detection and $t_{2}=1$~s for the Magnetoelectric Coupling direct detection.  
	
The influence of material volume on detection sensitivity is a critical consideration. For the first detection method (achieved through magnetization measurement), as previously mentioned, increasing the sample volume can further enhance sensitivity. However, in the second method (direct magnetoelectric coupling measurement), the dominant role of the intrinsic magnetoelectric coupling ($\theta_{\text{ME}}$) outweighs the axion-induced coupling ($\theta$). On one hand, enhancing the material's intrinsic coupling strength can effectively improve performance. On the other hand, since the measured voltage exhibits thickness dependence, increasing the thickness can amplify the statistically measured voltage values, thereby reducing measurement errors. Future work could explore stacking multiple film layers to effectively increase the detection volume while maintaining the unique strain-gradient properties.

In the context of axion detection, "amplification" refers to the process where the extremely weak signals generated by axion-electromagnetic field interactions are enhanced to detectable levels through the unique physical mechanisms of magnetoelectric materials. For Method 1 (achieved through magnetization measurement), the amplification factor can be quantitatively described by Equation \ref{eq:deltaM1}, which characterizes the conversion of weak axion signals into measurable magnetic responses.Indeed, it is precisely in the resonant regime (around meV) that nonlinear amplification becomes non-negligible. However, for most of the axion parameter space we are investigating, the detection frequencies are far from the resonant frequency, making linear amplification the dominant effect. We plan to conduct more precise investigations of amplification at resonant frequencies in future work. For Method 2 (direct magnetoelectric coupling measurement), the coupling occurs linearly between the axion field \(\theta\) and the material's magnetoelectric coupling \(\theta_{\text{ME}}\). Whether resonance occurs depends on the frequency characteristics of the \(\theta_{\text{ME}}\) field. If nonlinear components exist, Equation \ref{eq:theta1} would need to be extended to higher-order terms. The study of nonlinear \(\theta\) coupling represents an important direction for both future experimental and theoretical research.
	
	\section{Conclusion and Outlook}
	This study presents a novel approach for detecting axion dark matter using magnetoelectric (ME) materials, combining the strengths of SQUID-based magnetization measurements and direct magnetoelectric coupling experiments. The synergy between these methods not only enhances detection sensitivity but also provides a versatile platform for exploring axion interactions across different mass ranges. As a unique physical phenomenon, the magnetoelectric effect holds significant potential for novel dark matter detection techniques. Unlike traditional methods that require strong external magnetic fields, our approach leverages spontaneous symmetry breaking to promote the coupling between the axion field and the material's internal ferroelectric and magnetic orders. The material exhibits parallel magnetization and polarization, particularly when magnetic and polar phases coexist at the same temperature. This characteristic enhances the axion-electron coupling or the axion-photon-photon coupling when detecting the extremely weak axion dark matter particle, allowing for a more efficient conversion of the weak oscillation signals induced by the axion field into observable electric polarization or magnetization signals. Furthermore, resonance occurs when the magnon frequency matches the axion frequency, providing a promising and novel pathway for direct axion detection.
	
	We have measured and calculated $\text{Sr}_2\text{IrO}_4$ thin films with magnetoelectric coupling effects at low temperatures, deriving exclusion limits for axion-photon-photon coupling and axion-electron coupling. These limits can be significantly enhanced when the sample volume is sufficiently large and the measurement duration is extended. Our magnetoelectric coupling measurement system is also highly promising. This direct method for measuring $\theta$ is particularly convenient for studying axion-like phenomena in materials, and upon enhancing its precision, it will facilitate axion detection. In future work, we aim to improve detection sensitivity by exploring additional single-phase and composite magnetoelectric materials, including heterojunctions and superlattice designs. Under specific conditions, these materials could exhibit strong magnetoelectric coupling effects, enhancing their tunability, response frequency, and adjustment range. Additionally, we have observed the presence of excitation modes, including phonons, magnons, and ferrons, in magnetoelectric materials. Since these quasiparticles can be tuned by artificially designing their symmetry, it is possible to select specific modes in future axion detection research, which may provide additional opportunities for detecting dark matter through axion-matter coupling.

	% Acknowledgements
	\medskip
	\textbf{Acknowledgements} \par %delete if not applicable))
	This work is supported by the National Natural Science Foundation of China (No. 52225205, No. T2350005, No. 12105013 and No. 12447105), the National Key Research and Development Program of China (No. 2023YFA1406500 and No. 2021YFA0718700), the Fundamental Research Funds for the Central Universities and the Beijing Natural Science Foundation (Z240008).

	\medskip
	
	% Use the following code if you wish to generate your bibliography with BibTeX;
	% replace the string "MSP-template" below with the name(s) of
	% the BibTeX data base(s) you want to use.
	% The resulting bibliography-output (the content of the .bbl file)
	% must be pasted back into this file before submission.
	% Please also include your BibTeX data base file(s) in your submission
	% so that we can re-run BibTeX if necessary.
	%
	%\bibliographystyle{MSP}
	%\bibliography{MSP-template}
	
	%\textbf{References}\\
	
	\bibliographystyle {MSP}
    \bibliography{MSP-template}

	% Figures/tables and captions
	% Permission statements are required for all figures reproduced or adapted from previously published articles/sources. Please also ensure that all necessary permissions to reproduce images have been received
	% Please remove these statements for original figures

	\clearpage

	%\begin{figure}[t]
	%	\begin{center}
		%		\begin{minipage}[c]{0.5\textwidth}
			%			\includegraphics[height=6cm,width=8cm]{MER.png}
			%		\end{minipage}%
		%		\begin{minipage}[c]{0.5\textwidth}
			%			\includegraphics[height=6cm,width=8cm]{MM.png}
			%		\end{minipage}
		%		\caption{Time-dependent variation curves of magnetoelectric coupling strength and magnetization intensity.} \label{fig:MERMM1}
		%	\end{center}
	%\end{figure}

	% Please provide Biographies and photos for Essays, Feature Articles, Progress Reports, Reviews, and Perspectives for those authors who should be highlighted  
	% These should be at most 100 words long
	% For other article types this section can be removed
	% Photographs should be 40mm broad and 50 mm high

	% Table of contents entry should be 50 - 60 words long
	% Image should be 55 mm broad and 50 mm high or 110 mm broad and 20 mm high

\end{document}